\documentclass[aps,prl,twocolumn,showpacs,superscriptaddress,nofootinbib]{revtex4}
\usepackage{natbib}
\usepackage{epsfig}
\usepackage{graphicx}
\usepackage{float}
\usepackage{amsmath}
\usepackage{color}
\usepackage{amssymb}
\usepackage{amsfonts}
\usepackage{units}
\usepackage{dcolumn}
\usepackage{bm}
\usepackage[colorlinks,linkcolor=blue,anchorcolor=green,citecolor=blue]{hyperref}

\begin{document}

\title{Testing Einstein's weak equivalence principle with a 0.4-nanosecond giant pulse of the Crab pulsar}

\author{Yuan-Pei Yang}
\affiliation{Kavli Institute of Astronomy and Astrophysics, Peking University, Beijing 100871, China; yypspore@gmail.com}
\author{Bing Zhang$^{1,}$}
\affiliation{Department of Astronomy, School of Physics, Peking University, Beijing 100871, China}
\affiliation{Department of Physics and Astronomy, University of Nevada, Las Vegas, NV 89154, USA; zhang@physics.unlv.edu}

\date{\today}


\begin{abstract}
Einstein's weak equivalence principle (EEP) can be tested through the arrival time delay between photons with different frequencies. Assuming that the arrival time delay is solely caused by the gravitational potential of the Milky Way, we show that a ``nano-shot'' giant pulse with a time delay between energies
corrected for all known effects, e.g. $\Delta t<0.4~\unit{ns}$, from the Crab pulsar poses a new upper limit on the deviation from EEP, i.e. $\Delta\gamma < (0.6-1.8)\times 10^{-15}$. This result provides the hitherto most stringent constraint on the EEP, 
improving by at least 2 to 3 orders of magnitude from the previous results based on fast radio bursts.
\end{abstract}

\maketitle

Einstein's weak equivalence principle (EEP) is the foundation of general relativity. It states that the trajectories of any freely falling, uncharged test bodies are independent of their energy, internal structure, or composition.
The validity of general relativity and EEP can be characterized by the parametrized post-Newtonian (PPN) parameter, $\gamma$, which reflects the level of space curvature per by unit rest mass \cite{wil06}. For general relativity, $\gamma=1$ is predicted. The
measurement of the absolute value of $\gamma$ has reached high accuracy. The authors of Ref.\cite{lam09} derived $\gamma-1= (-0.8\pm1.2)\times10^{-4}$ via light deflection through very-long-baseline radio interferometry. The authors of Ref.\cite{ber03} obtained a more stringent constraint $\gamma-1=(2.1\pm 2.3)\times10^{-5}$ via the travel time delay of a radar signal from the Cassini spacecraft. On the other hand, in order to test the EEP, some people applied the arrival time delay to measure the difference of the $\gamma$ values for photons with different energies, or for different species of cosmic messengers (photons, neutrinos, and gravitational waves) \cite{gao15,wei15,tin16,nus16,sha64,li16}.

Adopting the PPN approximation, Shapiro time delay in a gravitational potential $U(r)$ is given by \cite{sha64}
\begin{eqnarray}
\delta t_{\rm{gra}}=-\frac{1+\gamma}{c^3}\int_{r_o}^{r_e}U(r)dr,\label{sd}
\end{eqnarray}
where $r_o$ and $r_e$ denote the locations of the observer and the emitting source, respectively. The relative Shapiro time delay $\Delta t_{\rm{gra}}$ between two photons with energy $E_1$ and $E_2$ caused by the gravitational potential $U(r)$ is then given by
\begin{eqnarray}
\Delta t_{\rm{gra}}=\frac{\gamma_1-\gamma_2}{c^3}\int_{r_o}^{r_e}U(r)dr.\label{eep}
\end{eqnarray}

Recently, Wei et al. \cite{wei15} applied this method to fast radio bursts (FRBs) passing through the gravitational field of the Milky Way (WM) galaxy and obtained a stringent constraint $\Delta \gamma<2.5\times10^{-8}$. Considering the large-scale gravitational potential fluctuations, an even more stringent constraint, $\Delta\gamma<4.5\times10^{-11}~(3\sigma)$ or $\Delta\gamma<2.8\times10^{-12}~(2\sigma)$, can be achieved \cite{nus16}. Based on a located fast radio burst FRB 150418 at $z=0.492\pm0.008$ \cite{kea16} (for a different view, see \cite{wil16}), Tingay \& Kaplan \cite{tin16} obtained $\Delta \gamma<(1-2)\times10^{-9}$ by only including the MW gravitational potential. When the large-scale gravitational potential fluctuations are included, a more stringent constraint, $\Delta\gamma<2.4\times10^{-12}~(3\sigma)$ or $\Delta\gamma<1.4\times10^{-13}~(2\sigma)$, is obtained \cite{nus16}. In addition to FRBs, some other extragalactic transient events, such as supernovae, gamma-ray bursts, active galactic nuclei, gravitational-wave sources, are also used to test EEP with this method, and the constraints are in the range of $\Delta\gamma<10^{-3}-10^{-10}$ \cite{gao15,li16}.

Here we show that the nanoseconds-long giant pulses observed from the Crab pulsar and several other pulsars can provide much more stringent constraints on $\Delta\gamma$. Even though these events are in the Milky Way galaxy, the effect of the extreme short duration, e.g. $\Delta t_{\rm{obs}}-\Delta t_{\rm{DM}}<0.4~\unit{ns}$ for one Crab pulsar giant pulse, overcompensates the deficit in distance, and leads to much more stringent constraint than extra-galactic ms-duration FRBs.

Giant pulses are one of the most striking phenomena of radio pulsars. They are characterized by extremely high fluxes (exceeding MJy \cite{sog07}) and ultra short durations. The typical duration is a few microseconds, but occasional bursts shorter than one nanosecond, the so-called ``nano-shots'', have been observed \cite{han07}. So far, they have been detected from the Crab pulsar \cite{sta68}, PSR B1937+21 \cite{sog04}, and some other pulsars \cite{rom01}.

The giant pulses from the Crab pulsar have been detected over a broad frequency range \cite{mof96}. Hankins \& Eilek \cite{han07} reported a giant pulse that showed a single, extremely intense pulse with flux exceeding 2 MJy and an unresolved dedispersed duration, $\Delta t_{\rm{obs}}-\Delta t_{\rm{DM}}<0.4~\unit{ns}$ \cite{han07}. The pulse was observed in a frequency band centered at 9.25 GHz with a 2.2 GHz bandwidth. 

In general, the observed time delay of two photons with different frequencies consists of five terms \cite{gao15,wei15}
\begin{eqnarray}
\Delta t_{\rm{obs}}=\Delta t_{\rm{int}}+\Delta t_{\rm{LIV}}+\Delta t_{m_\gamma}+\Delta t_{\rm{gra}}+\Delta t_{\rm{DM}},
\end{eqnarray}
where $\Delta t_{\rm{int}}$ is the intrinsic delay time that depends on the radiation mechanism of the source, $\Delta t_{\rm{LIV}}$ is the delay time caused by Lorentz invariance violation (LIV), $\Delta t_{m_\gamma}$ is the delay time due to nonzero photon rest mass \cite{wu16}, and $\Delta t_{\rm{DM}}$ is the delay time due to the cold plasma dispersion from free electrons along the line of sight, which is already corrected for radio pulsar pulse observations. 
If we assume that  $\Delta t_{\rm int}+\Delta t_{\rm{LIV}}+\Delta t_{m_\gamma}>0$, then $\Delta t_{\rm{gra}}=(\Delta t_{\rm{obs}}-\Delta t_{\rm{DM}})-(\Delta t_{\rm int}+\Delta t_{\rm{LIV}}+\Delta t_{m_\gamma})<\Delta t_{\rm{obs}}-\Delta t_{\rm{DM}}$. Thus, $\Delta t_{\rm{obs}}-\Delta t_{\rm{DM}}$ is the upper limit of $\Delta t_{\rm{gra}}$.
 
The Crab pulsar is located at $l=184.56^\circ$ and $b=-5.78^\circ$ \cite{man05}, which means that the Galactic center,  Earth, and the Crab pulsar are almost aligned in a straight line. The distance of Crab from Earth is $d=2.0~\unit{kpc}$ \cite{man05}, thus its distance to the Galactic center is approximately $r_{\rm{Crab}}\simeq d+s=10.3~\unit{kpc}$, where $s=8.3~\unit{kpc}$ is the distance of Earth from the Galactic center \cite{gil09}. The gravitational potential along the propagation path has contributions from the Milky Way and the Crab pulsar, i.e. $U(r)=U_{\rm{MW}}(r)+U_{\rm{Crab}}(r)$. The former can be estimated as $U_{\rm{MW}}(r)\simeq-(1-3)\times10^{15}~\unit{cm^2s^{-2}}$ at $r\simeq8-11~\unit{kpc}$ \cite{irr13}. 
The gravitational potential of the Crab pulsar can be estimated as $U_{\rm{Crab}}(x)=-GM/x$, where $M=1.4M_\odot$, and $x$ is the distance to the pulsar. Therefore one has
\begin{eqnarray}
\Delta\gamma&\equiv&|\gamma_1(10.35~\unit{GHz})-\gamma_2(8.15~\unit{GHz})|\nonumber\\
&<&c^3(\Delta t_{\rm{obs}}-\Delta t_{\rm{DM}})\left(|U_{\rm{MW}}|d+GM\ln\frac{d}{x_e}\right)^{-1} \nonumber\\
&\simeq&c^3(\Delta t_{\rm{obs}}-\Delta t_{\rm{DM}})/|U_{\rm{MW}}|d \nonumber\\
&\simeq& (0.6-1.8)\times10^{-15}.
\end{eqnarray}
Here we have made use of the fact $|U_{\rm{MW}}|d\gg GM\ln(d/R)$ and $x_e>R$, where $x_e$ is the radius where the giant pulse emission is emitted, and $R=10^6~\unit{cm}$ is the radius of the Crab pulsar. 

This result provides the hitherto most stringent constraint on the EEP.
Compared with the previous result derived from FRB 150418, even with the large-scale gravitational potential fluctuations taken into account, our result improves on the previous limits \cite{tin16,nus16} by at least 2 to 3 orders of magnitude.

\acknowledgments
We thank the anonymous referees for detailed suggestions that have allowed us to improve this manuscript significantly. This work is partially supported by the Initiative Postdocs Supporting Program (Grant BX201600003), Project funded by China Postdoctoral Science Foundation (Grant 2016M600851) and the National Basic Research Program (973 Program) of China (Grant 2014CB845800).

\bibliographystyle{apsrev4-1}

\begin{thebibliography}{99}

\bibitem{wil06} C.~M.~Will,  Living Rev. Relativ. {\bf 9}, 3 (2006);
C.~M.~Will,  Living Rev. Relativ. {\bf 17}, 4 (2014).

\bibitem{lam09} S. B. Lambert and C. Le Poncin-Lafitte, Astron. Astro-
phys. {\bf 499}, 331 (2009);
S. B. Lambert and C. Le Poncin-Lafitte, Astron. Astro-
phys. {\bf 529}, A70 (2011).

\bibitem{ber03} B. Bertotti, L. Iess, and P. Tortora, Nature (London) {\bf 425}, 374 (2003)

\bibitem{gao15} H. Gao, X.-F. Wu, and M\'{e}sz\'{a}ros, Astrophys. J. {\bf 810},
121 (2015).

\bibitem{wei15} J.-J. Wei, H. Gao, X.-F. Wu, and P. M\'{e}sz\'{a}ros, Physical
Review Letters {\bf 115}, 261101 (2015).

\bibitem{tin16} S. J. Tingay and D. L. Kaplan, Astrophys. J. Lett. {\bf 820}, L31 (2016).

\bibitem{nus16} A. Nusser, Astrophys. J. Lett. {\bf 821}, L2 (2016).

\bibitem{sha64} I. I. Shapiro, Physical Review Letters {\bf 13}, 789 (1964);
L. M. Krauss and S. Tremaine, Physical Review Letters {\bf 60}, 176 (1988);
M. J. Longo, Physical Review Letters {\bf 60}, 173 (1988).

\bibitem{li16} X. Li, Y.-M. Hu, Y.-Z. Fan, and D.-M. Wei, Astrophys. J. {\bf 827}, 75 (2016);
Y. Sang, H.-N. Lin, and Z. Chang, Mon. Not. R. Astron. Soc. {\bf 460}, 2282 (2016);
Z.-Y. Wang, R.-Y. Liu, and X.-Y. Wang, Physical Re- view Letters {\bf 116}, 151101 (2016);
J.-J. Wei, J.-S. Wang, H. Gao, and X.-F. Wu, Astrophys. J. Lett. {\bf 818}, L2 (2016);
X.-F. Wu, H. Gao, J.-J. Wei, P. M\'{e}sz\'{a}ros, B. Zhang, Z.- G. Dai, S.-N. Zhang, and Z.-H. Zhu, Phys. Rev. D {\bf 94}, 024061 (2016);
S.-N. Zhang, ArXiv e-prints (2016), arXiv:1601.04558.

\bibitem{kea16} E. F. Keane, S. Johnston, S. Bhandari, E. Barr, N. D. R.
Bhat, M. Burgay, M. Caleb, C. Flynn, A. Jameson, M. Kramer, E. Petroff, A. Possenti, W. van Straten, M. Bailes, S. Burke-Spolaor, R. P. Eatough, B. W. Stap- pers, T. Totani, M. Honma, H. Furusawa, T. Hattori, T. Morokuma, Y. Niino, H. Sugai, T. Terai, N. Tominaga, S. Yamasaki, N. Yasuda, R. Allen, J. Cooke, J. Jencson, M. M. Kasliwal, D. L. Kaplan, S. J. Tin- gay, A. Williams, R. Wayth, P. Chandra, D. Perrodin, M. Berezina, M. Mickaliger, and C. Bassa, Nature (London) {\bf 530}, 453 (2016). 

\bibitem{wil16} P. K. G. Williams and E. Berger, Astrophys. J. Lett. {\bf 821}, L22 (2016).

\bibitem{sog07} V. Soglasnov, in {\it WE-Heraeus Seminar on Neutron Stars
and Pulsars 40 years after the Discovery}, edited by
W. Becker and H. H. Huang (2007) p. 68.

\bibitem{han07} T. H. Hankins and J. A. Eilek, Astrophys. J. {\bf 670}, 693
(2007).

\bibitem{sta68} D. H. Staelin and E. C. Reifenstein, III, Science {\bf 162},
1481 (1968).

\bibitem{sog04} V. A. Soglasnov, M. V. Popov, N. Bartel, W. Cannon,
A. Y. Novikov, V. I. Kondratiev, and V. I. Altunin,
Astrophys. J. {\bf 616}, 439 (2004).

\bibitem{rom01} R. W. Romani and S. Johnston, Astrophys. J. Lett. {\bf 557},
L93 (2001);
S. Johnston and R. W. Romani, Astrophys. J. Lett. {\bf 590},
L95 (2003);
B. C. Joshi, M. Kramer, A. G. Lyne, M. A. McLaughlin,
and I. H. Stairs, in {\it Young Neutron Stars and Their Environments}, IAU Symposium, Vol. 218, edited by F. Camilo and B. M. Gaensler (2004) p. 319;
H. S. Knight, M. Bailes, R. N. Manchester, and S. M. Ord, Astrophys. J. {\bf 625}, 951 (2005).

\bibitem{mof96} D. A. Moffett and T. H. Hankins, Astrophys. J. {\bf 468}, 779 (1996);
J. M. Cordes, N. D. R. Bhat, T. H. Hankins, M. A. McLaughlin, and J. Kern, Astrophys. J. {\bf 612}, 375 (2004);
A. Jessner, A. S lowikowska, B. Klein, H. Lesch, C. H. Jaroschek, G. Kanbach, and T. H. Hankins, Advances in Space Research {\bf 35}, 1166 (2005);
A. Jessner, M. V. Popov, V. I. Kondratiev, Y. Y. Ko- valev, D. Graham, A. Zensus, V. A. Soglasnov, A. V. Bilous, and O. A. Moshkina, Astron. Astrophys. {\bf 524}, A60 (2010);
T. Eftekhari, K. Stovall, J. Dowell, F. K. Schinzel, and G. B. Taylor, ArXiv e-prints (2016), arXiv:1607.08612.

\bibitem{wu16} X.-F. Wu, S.-B. Zhang, H. Gao, J.-J. Wei, Y.-C. Zou, W.- H. Lei, B. Zhang, Z.-G. Dai, and P. M\'{e}sz\'{a}ros, Astrophys.
J. Lett. {\bf 822}, L15 (2016).



\bibitem{man05} J. M. Cordes and T. J. W. Lazio, ArXiv e-prints (2003), arXiv:0301598;
R. N. Manchester, G. B. Hobbs, A. Teoh, and M. Hobbs, Astron. J. {\bf 129}, 1993 (2005).

\bibitem{gil09} S. Gillessen, F. Eisenhauer, S. Trippe, T. Alexander,
R. Genzel, F. Martins, and T. Ott, Astrophys. J. {\bf 692},
1075 (2009).

\bibitem{irr13} A. Irrgang, B. Wilcox, E. Tucker, and L. Schiefelbein,
Astron. Astrophys. {\bf 549}, A137 (2013).

\end{thebibliography}

\end{document}